\documentclass[
	a4paper, %
	10pt, %
	unnumberedsections, %
	twoside, %
]{LTJournalArticle}

\runninghead{} %

\footertext{\textit{RISC-V Summit Europe, Bologna, 8-12th June 2026}} %

\usepackage{lipsum}
\usepackage[acronym]{glossaries}
\usepackage[detect-all=true,per-mode=symbol,per-symbol =/]{siunitx}
\usepackage[capitalise]{cleveref}
\usepackage{tablefootnote}
\usepackage{threeparttable}
\usepackage{booktabs}
\usepackage{xcolor}
\usepackage{makecell}
\usepackage{subcaption}

\DeclareSIUnit{\x}{\!\ensuremath{\times}}
\DeclareSIUnit\bit{b}
\DeclareSIUnit\gateeq{GE}
\glsenableentrycount
\let\gls\cgls
\let\glspl\cglspl
\let\Gls\cGls

\newacronym{asic}{ASIC}{application-specific integrated circuit}
\newacronym{dma}{DMA}{direct memory access}
\newacronym{eda}{EDA}{electronic design automation}
\newacronym{fpga}{FPGA}{field-programmable gate array}
\newacronym{ip}{IP}{intellectual property}
\newacronym{isa}{ISA}{instruction set architecture}
\newacronym{ml}{ML}{machine learning}
\newacronym{mcu}{MCU}{microcontroller}
\newacronym{obi}{OBI}{Open Bus Interface}
\newacronym{os}{OS}{open-source}
\newacronym{pdk}{PDK}{process design kit}
\newacronym{rtl}{RTL}{register transfer level}
\newacronym{soc}{SoC}{system-on-chip}
\newacronym{sv}{SV}{SystemVerilog}
\newacronym{vlsi}{VLSI}{very-large-scale integration}

\newcommand{\x}{$\times$}
\newcommand{\riscv}{\mbox{RISC-V}}

\title{HyperCroc: End-to-End Open-Source RISC-V MCU with a Plug-In Interface for\\Domain-Specific Accelerators} %

\author{%
    Philippe Sauter\textsuperscript{1}\thanks{Corresponding author: \href{mailto:phsauter@iis.ee.ethz.ch}{\tt phsauter@iis.ee.ethz.ch}}, \
    Thomas Benz\textsuperscript{1,2}, \
    Paul Scheffler\textsuperscript{1}, \
    Luca Benini\textsuperscript{1,3}
}

\date{
    \footnotesize\textsuperscript{\textbf{1}}Integrated Systems Laboratory, ETH Zurich \\
    \textsuperscript{\textbf{2}}~lowRISC C.I.C., Cambridge, United Kingdom\\
    \textsuperscript{\textbf{3}}Department of Electrical, Electronic, and
    Information Engineering, University of Bologna
}

\renewcommand{\maketitlehookd}{%
	\begin{abstract}
        \noindent Domain-Specific architectures with accelerators for machine learning and signal processing require efficient bulk data movement and high-bandwidth access to large datasets.
        Such capabilities are often absent from minimal open-source \glspl{mcu}.
        We present \emph{HyperCroc}, an extension to the end-to-end open-source \riscv{} Croc \gls{soc} integrating a silicon-proven \emph{HyperBus} controller for off-chip DRAM and Flash memory access and a DMA engine, providing a practical \gls{mcu}-class platform with streamlined plug-in support for domain-specific acceleration. 
        HyperBus offers a low-pin-count PSDRAM interface at up to \SI{400}{\mega\byte\per\second}, enabling bandwidth-scaled dataset access, while the DMA engine enables autonomous, high-throughput transfers without CPU intervention. 
        HyperCroc preserves Croc’s open-source synthesis and physical implementation flow targeting IHP’s open 130 nm \gls{pdk}; the full chip can be implemented in under one hour on a consumer-grade workstation. 
        We further report first silicon measurements from \emph{MLEM}, the first Croc tapeout, confirming that the silicon is fully functional at \SI{72}{\mega\hertz} @ \SI{1.2}{\volt} and validating the end-to-end flow.
	\end{abstract}
}

\begin{document}

\maketitle %

\glsresetall

\section{Introduction}

Access to semiconductor technology, \gls{eda} tools and reusable \gls{ip} blocks is a prerequisite to educate new chip designers and turn research prototypes into credible hardware. 
In the context of worldwide ``Chips Acts'' \cite{2024europeanchipsact,2024chipsandscienceact} and a persistent skill shortage, end-to-end \gls{os} platforms are increasingly important to lower the barrier to entry by making the entire design flow accessible.
The \gls{os} \riscv{} Croc \gls{soc} was developed in this spirit~\cite{sauter2025croc}, pairing a minimal \gls{mcu} with a streamlined RTL-to-GDS flow targeting IHP’s open \SI{130}{\nano\metre} \gls{pdk}, supported by documentation and lecture material with hands-on exercises.

At the same time, many important accelerator and signal-processing workloads are dominated by bulk data movement and access to large datasets. 
\Gls{fpga}-based prototypes can access high-bandwidth memory, but may not reflect \gls{asic} performance~\cite{Prabakaran2024} and integration constraints~\cite{darvishi2025pipelining}.
Conversely, \gls{asic}-oriented \gls{os} \glspl{mcu} such as Croc omit external memory interfaces, limiting their applicability as an extensible platform for domain-specific accelerators.

To address this gap, we present \emph{HyperCroc}, an extension of the Croc end-to-end \gls{os} \riscv{} \gls{mcu} that adds the memory infrastructure required to support domain-specific accelerator plug-ins.
HyperCroc integrates a silicon-proven HyperBus controller~\cite{sauter2025basilisk}, providing a low-power and low-pin-count PSDRAM interface connecting up to \SI{256}{\mebi\byte} PSDRAM (\SI{400}{\mega\byte\per\second}) or \SI{2}{\gibi\byte} Flash (\SI{333}{\mega\byte\per\second}) per PHY, together with an iDMA~\cite{benz2023iDMA} engine that enables autonomous, high-throughput data movement between external memory, on-chip memory, and accelerators. 
At the same time, HyperCroc shares Croc’s end-to-end \gls{os} design flow, enabling a full RTL-to-GDS implementation flow in under one hour on a consumer-grade workstation.

In addition to the description of the design IPs, we report on the first Croc silicon measurements, which confirm functional silicon at \SI{72}{\mega\hertz} and validate the platform and flow's maturity.
We present the following contributions:
\begin{itemize}
    \item \emph{HyperCroc}, an extensible {\riscv} \gls{mcu} platform with plug-in support for domain-specific accelerators, based on the Croc \gls{soc} with a silicon-proven \emph{HyperBus} controller and \emph{iDMA} to support high-bandwidth access to large datasets.
    \item A proven end-to-end open-source flow (software and RTL-to-GDS) targeting IHP's open \SI{130}{\nano\metre} \gls{pdk}, scaling Croc to MCU-class systems with high-bandwidth external memory and streamlined support for accelerator integration.
    \item First silicon measurements from \emph{MLEM}, Croc's initial tapeout, validating functional silicon and the maturity of the platform.
\end{itemize}

\section{Architecture}
\begin{figure}[t]
    \vspace{-0.9em}
    \centering
    \includegraphics[width=\linewidth]{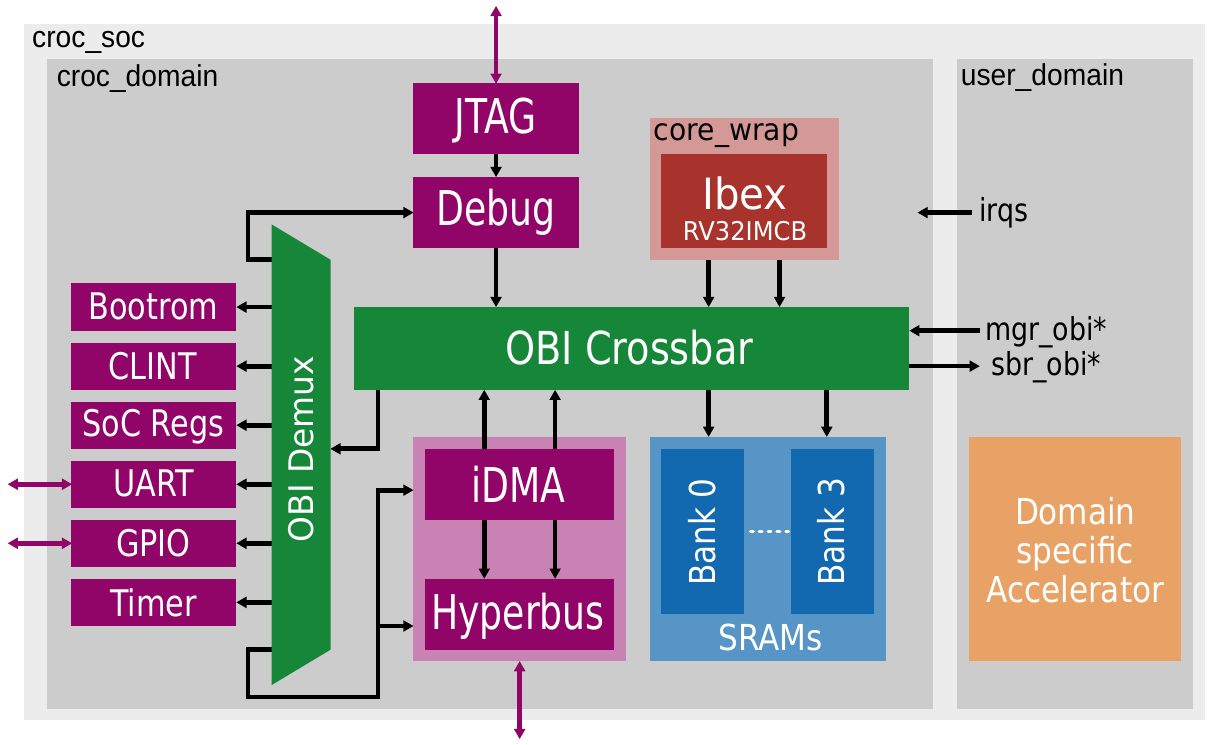}
    \vspace{-0.9em}
    \caption{HyperCroc architecture with Ibex core, iDMA and HyperBus controller.}
    \label{fig:hypercroc}
\end{figure}
Croc is a \gls{mcu}-class {\riscv} \gls{soc} organized into the \emph{Croc domain} and a plug-in interface (the \emph{user domain}) that allows domain-specific accelerators to be integrated with minimal effort.
The Croc domain combines a performance oriented 3-stage Ibex core (RV32IMCB), peripherals, and a tightly coupled \emph{single-cycle} interconnect to on-chip SRAM banks, achieving high per-cycle efficiency for \emph{control-heavy} workloads.

HyperCroc attaches the iDMA data mover to Croc’s main 32-bit \gls{obi} crossbar, giving it single-cycle access to on-chip SRAM. 
In addition, the iDMA connects directly to the HyperBus controller using \gls{obi} with burst support. 
Contiguous transactions are essential to minimize the overhead of the HyperBus protocol and reach peak sustained bandwidth.

Accelerators are attached to the main crossbar through the user domain and can rely on the iDMA + HyperBus path to off-chip memory or on-chip SRAMs for dataset ingress/egress. 
The HyperBus PHYs run in their own clock domain, up to \SI{200}{\mega\hertz}. Each can be connected to four HyperRAM/HyperFlash devices, providing up to \SI{2}{\gibi\byte} of storage per PHY. At \SI{200}{\mega\hertz} HyperBus reaches the maximum sustained throughput of \SI{400}{\mega\byte\per\second} for each PHY. With a typical \gls{soc} running at \SI{100}{\mega\hertz} or below, this is sufficient to transfer one 32-bit word per cycle. 

To simplify integration and ease timing closure, the HyperBus PHY is provided as a hardened macro: the high-speed signaling and timing-critical details are contained within the macro, leaving a clean, synchronous controller-facing interface for users.
For even higher bandwidth scenarios, HyperCroc can instantiate a dual-PHY controller to double external-memory capacity and bandwidth to \SI{2}{\gibi\byte} and \SI{800}{\mega\byte\per\second} while keeping the rest of the \gls{soc} unchanged.

\section{Croc Silicon Results}

We validated the baseline Croc \gls{mcu} by obtaining the first silicon measurements from MLEM, the first Croc-based demonstrator tapeout in IHP \SI{130}{\nano\metre}.
All peripherals, the core, and SRAM of the \emph{Croc domain} were fully verified using software-driven tests. The maximum operating frequency was measured and matches the expected sign-off results from the \gls{os} \gls{eda} tools.
This establishes the end-to-end \gls{os} \gls{mcu} platform and design flow as silicon-proven, reducing the integration and adoption risk for HyperCroc.

\cref{tab:metrics} summarizes baseline metrics obtained from silicon measurements and expected results for HyperCroc, highlighting the need for DMA in conjunction with external memory to build a capable system for accelerator development.

\begin{table}[t]
\centering
\begin{threeparttable}
\begin{tabular}{@{}lcc@{}}
\toprule
\textbf{Metric} & \textbf{Croc (silicon)} & \textbf{HyperCroc} \\
\midrule
Technology & \SI{130}{\nano\metre} & \SI{130}{\nano\metre} \\
Clock frequency & \SI{72}{\mega\hertz} @ \SI{1.2}{\volt} & \SI{100}{\mega\hertz} @ \SI{1.2}{\volt} \\
Internal Memory & 3$\times$\SI{8}{\kibi\byte} & 4$\times$\SI{8}{\kibi\byte} \\
External Memory & -- & 2\x\SI{256}{\mebi\byte} \\
Ext. bandwidth & -- & up to \SI{800}{\mega\byte\per\second} \\
\bottomrule
\end{tabular}
\begin{tablenotes}[flushleft]
\footnotesize
\item Croc values are obtained from MLEM chips measured on an
Advantest V93000 EXA Scale; HyperCroc values are design targets.
\end{tablenotes}
\end{threeparttable}
\caption{Croc and HyperCroc key metrics.}
\label{tab:metrics}
\end{table}

\section{Conclusion and Outlook}

HyperCroc extends the Croc end-to-end \gls{os} \riscv{} \gls{mcu} platform with iDMA and HyperBus, enabling realistic development and benchmarking of domain-specific accelerators, including \gls{ml} inference for embedded systems.
By strengthening the memory system while retaining a streamlined, reproducible \gls{os} RTL-to-GDS flow, HyperCroc bridges the gap between educational chip design and the research and prototyping of domain-specific accelerators while building on the same proven foundations.
HyperCroc will be taped out  March 2026 and released as open source, with silicon validation later this year.

\vspace{-0.2cm}

\printbibliography
\end{document}